\documentclass{iau}

\usepackage{amsmath}
\usepackage{graphicx}
\usepackage{multirow}

\newcommand {\ie} {{\it i.e.~}}

\newcommand {\eg} {{\it e.g.~}}

\newcommand{\msun}{M$_{\odot}$}

\newcommand{\Mstar}{\ifmmode M_{\star} \else $M_{\star}$\fi\xspace}
\newcommand{\MBH}{\ifmmode M_\text{BH} \else $M_\mathrm{BH}$\fi\xspace}
\newcommand{\Lrad}{\ifmmode L_{\rm 1.4} \else $L_{\rm 1.4}$\fi\xspace}
\newcommand{\Frad}{\ifmmode F_{\rm 1.4} \else $F_{\rm 1.4}$\fi\xspace}

\newcommand{\Ha}{H$\alpha$}
\newcommand{\Hb}{\ifmmode {\rm H}\beta \else H$\beta$\fi}

\newcommand{\hii}{H\,{\sc ii}}

\newcommand{\Nii}{[N\,{\sc ii}]\,$\lambda$6584}

\newcommand{\nii}{[N\,{\sc ii}]}

\newcommand{\oi}{[O\,{\sc i}]}

\newcommand{\oii}{[O\,{\sc ii}]}

\newcommand{\oiii}{[O\,{\sc iii}]}

\newcommand{\sii}{[S\,{\sc ii}]}

\begin{document}

\lefttitle{Cambridge Author}
\righttitle{Proceedings of the International Astronomical Union: \LaTeX\ Guidelines for~authors}

\jnlPage{1}{7}
\jnlDoiYr{2025}
\doival{10.1017/xxxxx}
\volno{395}
\pubYr{2025}
\journaltitle{Stellar populations in the Milky Way and beyond}

\aopheadtitle{Proceedings of the IAU Symposium}
\editors{J. Mel\'endez,  C. Chiappini, R. Schiavon \& M. Trevisan, eds.}

\title{Explaining the emission lines in early-type galaxies}

\author{Gra\.zyna Stasi\'nska}
\affiliation{LUX, Observatoire de Paris, Universit\'e PSL, Sorbonne Universit\'e, CNRS, 92190 Meudon, France (ORCID  0000-0002-4051-6146, e-mail: grazyna.stasinska@obspm.fr)}

\begin{abstract}
We explore the hypothesis that  the weak emission lines observed in some early-type galaxies (ETGs) are due to ionization by hot low-mass evolved stars (HOLMES) and analyze the \textit{pros} and \textit{cons.} 
\end{abstract}  


\maketitle

\section{Introduction}

Most early-type galaxies (ETGs) were once thought to be devoid of gas (see \eg a historic outline by Athey \& Bregman 2009). This was surprising  given that the total rate of mass-loss from red giants, planetary nebulae, supernovae, 
etc., is estimated to be of $0.1 - 1$ \msun\ per year (Mathews \& Baker 1971). In fact, as these authors and many others argued,  the gas produced by stellar mass-loss is heated by supernova explosions thereby giving rise to galactic winds, later observed as X-ray haloes around elliptical galaxies (White \& Davis 1998). 

With improving instrumentation,  several dedicated observational studies of ETGs (\eg Phillips et al. 1986) reported the existence of faint emission lines in about 50\% of the objects, this emission being extended and not concentrated around the nucleus in most cases. The Sloan Digital Sky Survey (York et al., 2000; Abazajian et al., 2009) has now obtained spectra of hundreds of thousands of galaxies and shown that about half of ETGs do present weak emission lines. Wide spectroscopic integral field surveys of galaxies such as the Calar Alto Legacy Integral Field Area survey (CALIFA) (S{\'a}nchez et al. 2012) or the Mapping nearby Galaxies at Apache Point Observatory  (MaNGA) survey (Bundy et al. 2015) have provided a wealth of data about the distribution of the intensities of these emission lines across the faces of the  galaxies, allowing deeper investigations regarding the origin of the gas and the ionizing mechanism producing these lines.

The proposed ionizing mechanisms are various, including ionization by post-AGB stars, active galactic nuclei, massive stars, internal shocks or electron conduction from hot gas, with the most favored one being post-AGB stars (Trinchieri \& di Serego Alighieri 1991, and many others).  Indeed, hot post-AGB stars are a natural explanation, since we know that they must be present in ETGs. The question is whether they are sufficient to explain the observed line intensities.

Binette et al. (1994) were the first to publish a quantitative study to explain emission-line luminosities and intensity ratios in ETGs due to post-AGB stars, using the evolutionary stellar population synthesis models of Bruzual \& Charlot (1993). In Section 2, we summarize the works based on the post-AGB iscenario. Note that since Cid-Fernandes et al. (2011) and Flores-Fajardo et al. (2011) the term post-AGB in this context has been replaced by hot low-mass evolved stars (HOLMES) to distinguish from ‘proto-planetary nebulae' which are commonly referred to as ‘post-AGBs’ in the planetary nebula community. The term HOLMES has also the advantage of including stars that may emit ionizing photons without having reached the tip of the AGB. Therefore we use it consistently in the remaining of this paper, even if the quoted original papers use the term post-AGB stars. A more extensive review of the role of HOLMES can be found in Stasi\'nska et al. (2022). 
In Section 3, we examine indices for the origin of the emitting gas. 
In the last section, we review arguments that have been advanced against the hypothesis of HOLMES to explain the emission lines in ETGs, showing that the problem has not yet been fully resolved.

\section{A short review of the HOLMES scenario to explain emission lines in ETGs}

Little after the publication of the Binette et al. (1994) paper, Macchetto et al. (1996) presented a survey of the ionized interstellar medium (ISM) in ETGs and concluded that indeed HOLMES can explain the observed equivalent widths and the spatial distribution of the emission lines.

With SDSS observations of galaxies, it is now possible to infer the characteristics of the stellar populations covered by the observing fiber (3 arcsec in diameter), and accurately measure the intensities of the emission lines observed in the same spectra after removing the stellar continuum (\eg Cid Fernandes et al. 2005). Using the characteristics of the stellar populations uncovered by such reverse population synthesis, Stasi\'nska et al. (2008) showed that many of the galaxies previously thought to be ionized by active galactic nuclei (AGNs) as well as those with  LINER-type spectra can actually be ionized by their own HOLMES. They called these objects `retired galaxies'. Compared to AGNs or true LINER-galaxies, these objects have very small equivalent widths of their emission lines, a property that cannot be read-out from the classical BPT (Baldwin et al. 1981) line-ratio diagram. In fact, true LINERS occupy roughly the same region as retired galaxies in the BPT diagram. This led Cid Fernandes et al. (2011) to propose a new diagram to distinguish retired galaxies, based on the equivalent width of \Hb\ and on the \Nii/\Ha\ ratio (the WHAN diagram).
\textit{Nota bene,} these studies have changed our view on the demographics of AGN galaxies, among the whole population of galaxies, AGN galaxies being less numerous than previously thought (see Cid Fernandes et al., 2010, Stasi\'nska et al. 2015, Stasi{\'n}ska et al. 2025)

Other studies by several other groups, sometimes in the context of LINERs or post-starburst galaxies,  also came to the conclusion that HOLMES play an important/dominant role in the ionization of ETGs (Taniguchi et al. 2000, Annibali et al. 2010, Capetti \& Baldi 2011, Yan \& Blanton 2012, Gutkin et al. 2016).

All the works mentioned above are based on  stellar population models using post-AGB evolution tracks computed by Vassiliadis \& Wood (1994) or  Bloecker (1995) and Schoenberner (1983). New  evolution models for post-AGB stars have been computed since then by Miller Bertolami (2016), which contain updated microphysics and include convective boundary mixing during the thermal pulses on the AGB. As a result, these models are at least three to ten times faster  that the previous ones, and they are also 0.3 dex brigthter. The effect of including these new evolutionary tracks in stellar population synthesis models has been investigated by Stasi\'nska et al. (2023). It turns out that the changes with respect to models using the old tracks are not dramatic. The models of Miller Bertolami (2016) have now been incorporated in the new version of the Bruzual \& Charlot (2003) stellar population models (see Mart{\'\i}nez-Paredes et al. 2023).

These last two decades the use of integral field spectroscopy (IFS) has allowed the study of the distribution of emission lines and comparison with the distribution of stellar light in galaxies, as well as the investigation of the detailed kinematics of these objects. 
For example, Sarzi et al. (2007) found that round and slowly rotating ETGs generally display uncorrelated stellar and gaseous angular momenta, whereas in flatter and fast rotating galaxies gas and stars are co-rotating.
Analyzing a sample of 32 ETGs from the CALIFA survey, Gomes et al. (2016) found that in half
of them the emission line ratios are LINER-type in both the nuclear and
extranuclear zones and that the distribution of the \Ha\ equivalent width is compatible with the HOLMES hypothesis. The
other half of the galaxies have no gas in the central zones. These authors also found that many of their galaxies
 show decoupling between gas and stellar kinematics.

This leads us to the intriguing question of the origin of gas producing the emission lines in ETGs.

\section{Where does the gas come from?}

The mass of the  gas originating from stellar mass-loss during the evolution of ETGs  is larger by orders of magnitudes than that needed to produce the observed emission lines (see \eg \ Herpich et al, 2018). However, as mentioned above,  the gas from mass-loss is heated to X-ray temperatures and does not emit in the optical (Mathews 1990, Bregman \& Parriott 2009, Li et al. 2019).

About half of the ETGs actually do not show emission lines at all. Yet, the HOLMES in  the lineless galaxies and those in late-type galaxies that have emission lines produce the same number of ionizing photons. This has been clearly demonstrated by Herpich et al. (2018), who pair-matched the two samples in stellar mass, redshift and half-light radius in the  r band. This suggests that the absence of emission lines is not due to a lack of ionizing photons, but to a lack of warm gas. Emission-line galaxies show a certain degree of dust extinction, which is absent in lineless galaxies. This again argues for an absence of gas in lineless galaxies. Why is it that some ETGs contain warm gas and others don't?

We already mentioned that in some galaxies the kinematics of the gas and of the stars are decoupled, suggesting an external origin for the gas.
In addition Herpich et al. (2018) found that in the galaxies exhibiting lines,  the \nii/\oii\ sequence as a function of galaxy mass
is the exact prolongation of the \nii/\oii\ sequence of star-forming galaxies, indicating that the emitting gas is
not enriched in nitrogen as it would be if it where originating from planetary nebulae. This is an additional argument for its external origin.
Herpich et al. (2018) also found traces of recent star formation in galaxies with emission lines, and suggested that 
these stars must have been made of gas coming probably from accretion from the haloes of the galaxies, or from residual
streams of metal-rich gas coming from a merger in the recent past. 

All the above converges to conclude that the ionization source and the origin of the gas producing the emission lines are disconnected!

\section{Some pending questions}

The so-called UV upturn observed in some bright ETGs, \ie the rise in their flux between 2500 and 1500 \AA\ is generally ascribed to old stellar populations (Yi 2008). If this explanation is correct, the UV images of these galaxies should be as smooth as optical images. However, examining a large sample of galaxies with UV upturn, Pandey et al. (2024) found that the clumpiness of ETGs is significantly larger 
in the UV than in the optical indicating that  the UV flux is dominated by young stars, which implies that star formation exists in all ETGs in the nearby Universe. If this is true, then the observed emission lines would be produced by young, massive stars. However, it would remain to explain why the observed emission-line spectra do not look like those of \hii\ regions but are of LINER-type.

Among the competing explanations for the excitation of emission lines in ETGs are interstellar shocks (Dopita \& Sutherland 1995). Apart from the worry to build a quantitative scenario for these shocks that would not be \textit{ad hoc}, the difficulty arises of characterizing shock emission. Shocks produce strong lines from low ionization species like \oii, \nii, \sii, and \oi. But this is also the case of X-ray ionization. As a matter of fact, it is partly the X-rays produced by the shocks which create this enhancement of low excitation lines, and the latter can also be produced in diluted \hii\ regions. A much clearer signature of shocks is the high electron temperatures produced by shock heating. Lee et al. (2024)  selected a sample of 2795 quiescent red-sequence galaxies from the MaNGA survey, and stacked their spectra to  measure the temperatures using different temperature-sensitive line ratios. They find that the temperatures are consistent with models of interstellar shocks (Alarie \& Morisset 2019) at solar metallicity and support shocks as the dominant ionization source in ETGs.
The most important indication comes from the \oiii\ line ratios which indicate much higher temperatures than the other ones. Note that high \oiii\ temperatures have also been measured with the same stacking technique in star-forming galaxies (Khoram \& Belfiore 2024) and these authors suggested that they might result from imperfections in the flux measurements or additional, yet undetermined physical processes. So, for the time being, the question of the \oiii\ temperature requires more work before being used as an indication of the dominance of shocks in ETG line emission.

Another argument that is being used against the hypothesis of  HOLMES being the powering agent of the emission lines in ETGs is the dearth of post-AGB detected with the \textit{Hubble Space Telescope} in the nearby elliptical galaxy M32 (Brown et al. 2008). The much faster evolution of post-AGB stars predicted by modern stellar models (Miller Bertolami 2016) implies that the number of  post-AGB stars is less than previously thought. However, as noted by this author, this is proably not sufficient to explain the discrepancy between the estimated number of post-AGB stars and the observed one. An additional factor, suggested by Brown et al. (2008) but dismissed by them, is that the post-AGB stars are enshrouded in circumstellar material. This option has been examined by Sarzi et al. (2011) in their study of the planetary nebula population in the central regions of M32.  They conclude that extinction could indeed contribute to the explanation of the lack of observed post-AGB stars in M32.

\end{document}